\documentclass[12pt]{iopart}
\usepackage{iopams}
\usepackage{graphicx}
\begin{document}
\jl{3}

\article[HFM2006]{Symposium paper}
{Magnetic Ordering of CoCl$_{2}$-GIC: a Spin Ceramic -Hierarchical Successive Transitions and the Intermediate Glassy Phase-}

\author{Masatsugu Suzuki\dag\footnote[3]{suzuki@binghamton.edu}, Itsuko S.
Suzuki\dag  , 
and Motohiro Matsuura\ddag\footnote[4]{matsuura@ccmails.fukui-ut.ac.jp}}

\address{\dag\ Department of Physics, State University of New York at Binghamton, Binghamton, New York 13902-6000, USA}

\address{\ddag\ Department of Management and Information Science, Fukui University of Technology, Fukui, Fukui 910-8505, JAPAN}

\begin{abstract}
Stage-2 CoCl$_{2}$-GIC is a spin ceramic and shows hierarchical successive transitions at $T_{cu}$ (= 8.9 K) and $T_{cl}$ (= 7.0 K) from the paramagnetic phase into an intra-cluster (two-dimensional ferromagnetic) order with inter-cluster disorder and then to an inter-cluster (three-dimensional antiferromagnetic like) order over the whole system. The nature of the inter-cluster disorder was suggested to be of spin glass by nonlinear magnetic response analyses around $T_{cu}$ and by studies on dynamical aspects of ordering between $T_{cu}$ and $T_{cl}$. Here, we present a further extensive examination of a series of time dependence of zero-field cooled magnetization $M_{ZFC}$ after the aging protocol below $T_{cu}$. The time dependence of the relaxation rates $S_{ZFC}(t) = (1/H)dM_{ZFC}(t)/d\ln t$ dramatically changes from the curves of simple spin glass aging effect below $T_{cl}$ to those of two peaks above $T_{cl}$. The characteristic relaxation behavior apparently indicates that there coexist two different kinds of glassy correlated regions below $T_{cu}$. 
\end{abstract}



\maketitle

\section{\label{intro}Introduction}
From the viewpoint of cooperative dynamics, a spin ceramic is an interesting system. It is composed of mesoscopic spin clusters on the regular lattice, coupled mutually through random interface boundaries. It is heterogeneous morphologically. The ordering characteristic should be different qualitatively from both in regular spin systems and in spin glasses. An attractive system is a stage-2 CoCl$_{2}$ graphite intercalation compound (GIC), in which each intercalate CoCl$_{2}$ layer is not extended infinitely, forming island-like ferromagnetic (F) clusters of mesoscopic scale. It shows successive magnetic transitions of a hierarchical nature at temperatures $T_{cu}$ (= 8.9 K) and $T_{cl}$ (= 7.0 K) and was identified to be from the paramagnetic into the three-dimensional (3D) antiferromagnetic (AF) order over the whole system through 2D ferromagnetic (F) order between $T_{cu}$ and $T_{cl}$ \cite{enoki,ref1,miyoshi}.

Such a picture, however, was not so appropriate because a number of peculiar phenomena observed so far suggested rather that the intermediate state was a disordered state among the clusters and also that the nature was of a spin glass \cite{enoki,ref1,miyoshi}. Especially, nonlinear magnetic response analysis in the vicinity of $T_{cu}$ and recent study of dynamical aspect of ordering processes established that the transition at $T_{cu}$ is certainly of a spin glass type \cite{ref1,miyoshi}. Surprisingly in the latter, the time dependence of the relaxation rates $S_{ZFC}(t)$ defined in section \ref{exp}, was found to have two distinguishable peaks in the temperature between $T_{cu}$ and $T_{cl}$, forming a contrast to the simple spin glass feature observed below $T_{cl}$ \cite{ref1}. Such relaxation phenomena apparently indicate that two correlated regions coexist in the intermediate temperatures, reflecting a complex situation of the intermediate glassy phase so far speculated. Under such a circumstance, we have carried out, in this work, a further extensive and detailed study of the aging effect on the relaxation processes of $M_{ZFC}(t)$ in a wider temperature region below $T_{cu}$.

\section{\label{exp}Experimental procedure}
\begin{figure}
\begin{center}
\includegraphics[width=8.0cm]{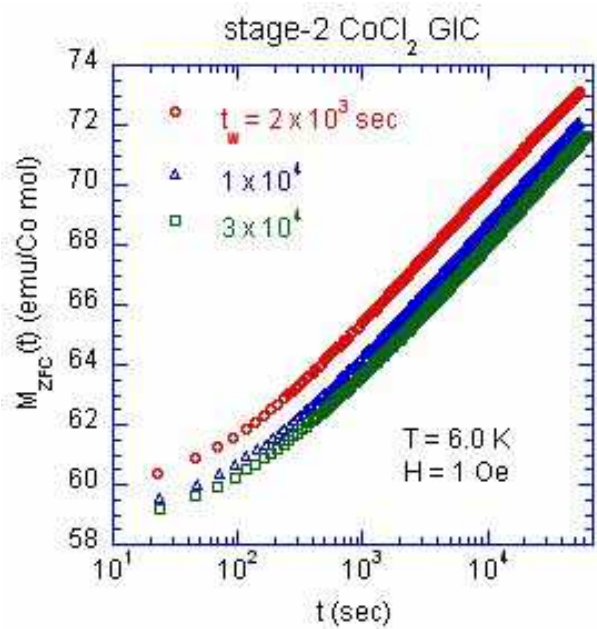}%
\end{center}
\caption{\label{fig1}$t$ dependence of $M_{ZFC}(t)$ at $T$ = 6.0 K for stage-2 CoCl$_{2}$ GIC. The system was cooled from 50 K at $T$ and isothermally aged at a wait time $t_{w}$ ($= 2.0\times 10^{3}$, $1.0\times 10^{4}$, $3.0\times 10^{4}$ sec). Immediately after the magnetic field is switched on to $H$ (= 1 Oe), the ZFC magnetization was measured as a function of $t$.}
\end{figure}

We have measured the time dependence of $M_{ZFC}(t)$ of stage-2 CoCl$_{2}$-GIC. Before $M_{ZFC}(t)$ measurement, the following ZFC aging protocol was taken. First the system was annealed at 50 K, a temperature far above $T_{cu}$ for $1.2\times 10^{3}$ sec. Then the system was quenched from 50 K to the measured temperature $T$ ($<T_{cu}$) in the absence of magnetic field and isothermally aged at the temperature for a wait time $t_{w}$ ($= 2.0\times 10^{3}$ sec, $1.0\times 10^{4}$ sec, $3.0\times 10^{4}$ sec in the present case). Immediately after the magnetic field $H$ (= 1 Oe) is turned on at $t$ = 0, $M_{ZFC}(t)$ was measured as a function of time $t$. Figure \ref{fig1} shows an example of the relaxation process of $M_{ZFC}$ plotted against $t$ for different $t_{w}$ at 6.0 K, a temperature well below $T_{cl}$. It clearly indicates an aging time dependence of $M_{ZFC}(t)$. The aging characteristic, however, could be demonstrated more distinguishably in the relaxation rate $S_{ZFC}(t)$, which is derived from $M_{ZFC}(t)$ as
\begin{equation}
S_{ZFC}(t) = (1/H)dM_{ZFC}(t)/d\ln t.
\label{eq02}
\end{equation}

\section{\label{result}Experimental result and discussion}

\begin{figure}
\begin{center}
\includegraphics[width=9.0cm]{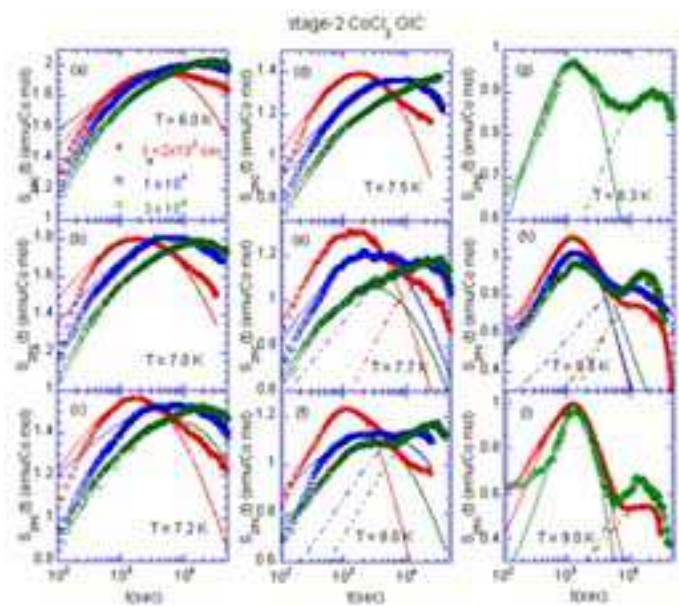}%
\end{center}
\caption{\label{fig2}$t$ dependence of the relaxation rate $S_{ZFC}(t)$. $t_{w} = 2.0\times 10^{3}$, $1.0\times 10^{4}$, and $3.0\times 10^{4}$ sec. The solid lines are curves fitted to the SER form. (a) $T$ = 5.5 K. (b) 6 K, (c) 7.0 K, (d) 7.2 K, (e) 7.5 K, (f) 7.7 K, (g) 8.0 K, (h) 8.5 K, and (i) 9.0 K.}
\end{figure}

\begin{figure}
\begin{center}
\includegraphics[width=8.0cm]{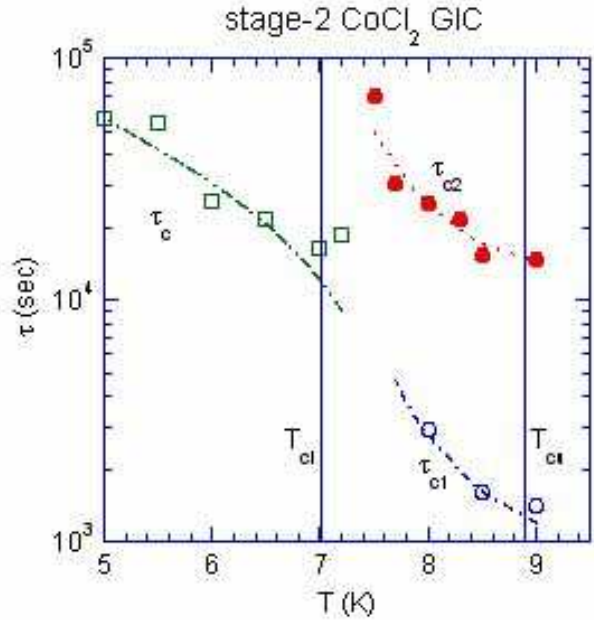}%
\end{center}
\caption{\label{fig3}$T$ dependence of the SER relaxation time $\tau$, which is derived from the least squares fit of $S_{ZFC}(t)$ vs $t$ ($H$ = 1 Oe, $t_{w} = 3.0\times 10^{4}$ sec) to the SER form. The SER relaxation time $\tau$ is almost equal to the peak time $t_{cr}$ of $S_{ZFC}(t)$ vs $t$ for $T<T_{cl}$ and 8.0 K $<T<T_{cu}$. Two SER peaks are observed at $t<3.0\times 10^{3}$ sec ($\opencircle$) and at $t>1.0\times 10^{4}$ sec ($\fullcircle$) above 8.0 K, while a single peak below $T_{cl}$($\opensquare$). The solid lines are guide to the eyes.}
\end{figure}

Figures \ref{fig2}(a)-(i) show the $S_{ZFC}(t)$ vs $t$ curves at various $t_{w}$ for $T$ = 6.0 -- 9.0 K. Figure \ref{fig2}(a) shows the $S_{ZFC}(t)$ vs $t$ curves for various $t_{w}$ at 6.0 K. Each curve exhibits a single broad maximum around $t_{w}$, and shifts to longer-$t$ side with increasing $t_{w}$. Such an aging time dependence of the relaxation rate is well known as a characteristic feature of spin glass dynamics and has been found actually in typical spin glass systems \cite{ref2}. Figure \ref{fig2}(h) shows the $S_{ZFC}(t)$ vs $t$ curves at 8.5 K, a temperature well above $T_{cl}$ and below $T_{cu}$. The $S_{ZFC}(t)$ vs $t$ curves show two separated peaks, forming a remarkable contrast to those at 6.0 K. Such a qualitative difference apparently indicates two different relaxation processes and therefore coexistence of two different correlated regions in the present system \cite{ref2}. The peaks appear at about $2.0\times 10^{4}$ sec and $1.0\times 10^{3}$ sec, respectively and the location of the peaks seems to be almost independent of $t_{w}$. In such a way, the relaxation behaviors for $T<T_{cl}$ and for $T>T_{cl}$ are qualitatively different. So, in order to examine how the characteristic relaxation behavior for $T>T_{cu}$ changes to that for $T<T_{cl}$ with decreasing $T$ and to identify the glassy state in the temperature range between $T_{cu}$ and $T_{cl}$, we observe the time dependence of $S_{ZFC}(t)$ systematically at various successive temperatures below $T_{cu}$.

As shown in Figs.~\ref{fig2}(a)-(i), $S_{ZFC}(t)$ vs $t$ curves for $T<T_{cl}$ exhibit a single broad peak around a characteristic time $\tau_{c}$, which shift to longer-$t$ side with increasing $t_{w}$. It reflects that the relaxation process and the aging effect are qualitatively the same as those at 6.0 K. Therefore we conclude that the state below $T_{cl}$ is of a typical spin glass. We also find that $S_{ZFC}(t)$ vs $t$ curves for well above $T_{cl}$ (8.0 K $<T<T_{cl}$) exhibit two peaks at $\tau_{c1}$ and $\tau_{c2}$ ($>\tau_{c1}$) suggesting the state in the region is essentially the same as that at 8.5 K. The values of $\tau_{c1}$ and $\tau_{c2}$ seems to be almost independent of $t_{w}$ (see Figs.~\ref{fig2}(h) and (i)). The $S_{ZFC}(t)$ vs $t$ curves above and near $T_{cl}$ ($T_{cl}<T<8.0$ K), however, are complicated and intermediate more or less. The identification of the state in this temperature region is thus difficult. So we tried to analyze all the data as a superposition of stretched exponential relaxation (SER) curves \cite{ref3}. The obtained temperature dependence of the SER relaxation time $\tau$ is summarized in Fig.~\ref{fig3}, where $\tau$ is equal to $\tau_{c}$ for $T<T_{cl}$ and separated into $\tau_{c1}$ and $\tau_{c2}$ for 8.0 K $<T<T_{cu}$. From the relaxation behavior and the characteristic aging effect in Figs.2 and 3, the system below $T_{cl}$ is apparently in the spin glass ordered state, while the system above $T_{cl}$ is roughly divided in two correlated domains or sub-systems. One is characterized by a SER process of longer relaxation time $\tau_{c2}$ and the other by another SER process of shorter relaxation time $\tau_{c1}$. As shown in Fig.~\ref{fig3}, the value of $\tau_{c2}$ is very long ($\approx 2\times 10^{4}$ sec) and increases with decreasing temperature. Below about 7.5 K, the relaxation process is not observable, probably due to the longer relaxation time. The value of $\tau_{c1}$ increases also as $T$ decreases.  The relaxation process of this $\tau_{c1}$ seems to change the character across $T_{cl}$ and to turn finally into the spin glass state below $T_{cl}$.  The relaxation process and its aging time dependence below $T_{cl}$ is in a reasonable agreement with the spin glass like memory phenomena of $M_{TRM}$ (thermoremanent magnetization) and $M_{ZFC}$ so far observed in a series of heating and cooling processes below $T_{cl}$ of the system\cite{ref1,ref6}.

\section{Concluding remarks}
From the present experimental result we conclude that the nature of the intermediate state between $T_{cu}$ and $T_{cl}$ is qualitatively different from that below $T_{cl}$, where aging dynamics of a typical spin glass was found.  In the intermediate state, the system is divided into two sub-systems. One goes into a spin glass like ordered state at $T_{cu}$. The other goes into a similar glassy state below $T_{cu}$ but of much shorter relaxation time. 

The latter sub-system is probably composed of the ferromagnetic moments ordered already below $T_{cu}$ within each intercalated cluster, referring to the previous experimental results. Then, the former may be speculated to consist of the spins located in the boundary region of each intercalated cluster and to affect on the latter as a random field because of the much longer relaxation time. Such a picture on the present system in the intermediate temperature between $T_{cu}$ and $T_{cl}$ is expected to bring a reasonable explanation of the remained questions including the anomalous memory phenomena of $M_{TRM}$ and $M_{ZFC}$ across $T_{cl}$\cite{ref1,ref6}.
 
\ack
We are grateful to Prof. H. Suematsu for providing us with single crystals of kish graphites. This work is supported in part by FUT Research Promotion Fund.

\end{document}